\documentclass[a4paper,11pt]{article}

\usepackage{jinstpub} 
\usepackage[binary-units]{siunitx}
\usepackage{subfigure}

\title{Performance of the large scale HV-CMOS pixel sensor MuPix8}

\author[a,1]{H.~Augustin,\note{Corresponding author.}}
\author[b]{N.~Berger,}
\author[a]{C.~Blattgerste,}
\author[a]{S.~Dittmeier,}
\author[c]{F.~Ehrler,}
\author[b]{C.~Grzesik,}
\author[a]{J.~Hammerich,}
\author[a]{A.~Herkert,}
\author[a,2]{L.~Huth,\note{Now at DESY, Hamburg, Germany}}
\author[a]{D.~Immig,}
\author[b]{A.~Kozlinskiy,}
\author[b]{M.~K\"oppel,}
\author[a,3]{J.~Kr\"oger,\note{Now at CERN, Switzerland}}
\author[a,4]{F.~Meier,\note{Now at PSI, Switzerland}}
\author[a]{A.~Meneses Gonzales,}
\author[b]{M.~M\"uller,}
\author[a]{L.~Noehte,}
\author[c]{I.~Peri\'{c},}
\author[c]{M.~Prathapan,}
\author[a]{T.~Rudzki,}
\author[c]{R.~Schimassek,}
\author[a]{A.~Sch\"oning,}
\author[b]{I.~Sorokin,}
\author[b]{F.~Stieler,}
\author[b]{A.~Tyukin,}
\author[b]{T.~Wagner,}
\author[b]{F.~Wauters,}
\author[a,c]{A.~Weber,}
\author[a,5]{D.~Wiedner,\note{Now at TU Dortmund, Germany}}
\author[c]{~H.~Zhang,}
\author[b]{and~M.~Zimmermann}

\affiliation[a]{Physikalisches Institut, Universit\"at Heidelberg,\\Im Neuenheimer Feld 226, 69120 Heidelberg, Germany}
\affiliation[b]{Institut f\"ur Kernphysik, Johannes Gutenberg-Universit\"at Mainz,\\Johann-Joachim-Becherweg 45, 55128 Mainz, Germany}
\affiliation[c]{Institut f\"ur Prozessdatenverarbeitung und Elektronik, KIT,\\Hermann-von-Helmholtz-Platz 1, 76344 Eggenstein-Leopoldshafen, Germany}

\emailAdd{augustin@physi.uni-heidelberg.de}

\abstract{
The Mu3e experiment is searching for the charged lepton flavour violating decay $ \mu^+\rightarrow e^+ e^- e^+ $, aiming for an ultimate sensitivity of one in $10^{16}$ decays. In an environment of up to $10^9$ muon decays per second the detector needs to provide precise vertex, time and momentum information to suppress accidental and physics background. The detector consists of cylindrical layers of $ 50\, \text{\textmu m} $ thin High Voltage Monolithic Active Pixel Sensors (HV-MAPS) placed in a $1\,\text{T}$ magnetic field. The measurement of the trajectories of the decay particles allows for a precise vertex and momentum reconstruction. Additional layers of fast scintillating fibre and tile detectors provide sub-nanosecond time resolution.
The MuPix8 chip is the first large scale prototype, proving the scalability of the HV-MAPS technology. It is produced in the AMS aH18 $180\, \text{nm}$ HV-CMOS process. It consists of three sub-matrices, each providing an untriggered datastream of more than $10\,\text{MHits}/\text{s}$. The latest results from laboratory and testbeam characterisation are presented, showing an excellent performance with efficiencies $>99.6\,\text{\%}$ and a time resolution better than $10\, \text{ns}$ achieved with time walk correction.
}


\proceeding{\normalsize 9$^{\text{th}}$ International Workshop on Semiconductor Pixel Detectors for Particles and Imaging\\
  10-14 December 2018\\
  Academia Sinica, Taipei}

\begin{document}
\maketitle
\flushbottom


\section{Introduction}
\label{sec:intro}

%
%

To achieve the goals of the Mu3e experiment~\cite{RP} an ultra-thin pixel tracker with $1\,\text{\textperthousand}$ of radiation length per layer is being built which detects up to \(10^9\) muon decays per second. Currently used tracking detector technologies can not fulfill both the material budget and high rate requirements.

The technology enabling to build this type of detector was found in the High Voltage Monolithic Active Pixel Sensors (HV-MAPS), see figure \ref{fig2}, which are produced in a commercially available HV-CMOS process~\cite{Peric:2007zz}. The combination of drift based charge collection and integrated readout on the same chip allows to build fast pixel detectors with an excellent fill-factor. Furthermore the small active depletion region allows the thinning of the sensors to $ 50\, \text{\textmu m} $ thickness.

A series of small prototypes, the so-called MuPix chips, have been designed and tested to show the feasibility of the HV-MAPS concept. The development culminated in the MuPix7 chip which is the first fully monolithic chip build in this technology and showed an excellent performance~\cite{Augustin:2016hzx,PIXEL:MuPix7,Augustin:2016pwd,Augustin:2017guc,Augustin:2018ppf}.
With the MuPix8 chip the scalability of the technology is tested.
In the following the architecture of the chip is described and key results of the laboratory and testbeam characterisation are presented.

\begin{table}[h]
\centering
\begin{tabular}{|l|cc|}
\hline
Chip & MuPix7 & MuPix8\\
\hline
Substrate resistivity & \(20\ \Omega\text{cm}\) & \(80\, \&\, 200\,\Omega\text{cm}\) \\
Chip size & \(0.35\times0.4\,\text{cm}^2\) & \(1\times2\,\text{cm}^2\)\\
Matrix size & \(40\times32\) pixels & \(128\times200\) pixels\\
Pixel size & \(103\times80\,\text{\textmu m}^2\) & \(81\times80\,\text{\textmu m}^2\) \\
\hline
\end{tabular}
\caption{A comparison of MuPix7 and MuPix8.}
\label{tab1}
\end{table}

\newpage
\section{The MuPix8 Chip}
\label{sec:mupix8}
%
%
%
%
%
%

\subsection{The Architecture}
\begin{figure}[h]
\centering
\subfigure[Sketch of the HV-MAPS principle~\cite{Peric:2007zz}.]{\includegraphics[width = .5\textwidth]{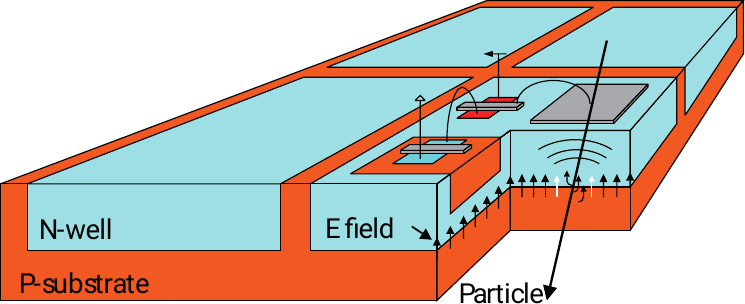}\label{fig2}}
\subfigure[The MuPix8 chip~\cite{Kroeger2017}.]{\includegraphics[height = .34\textheight]{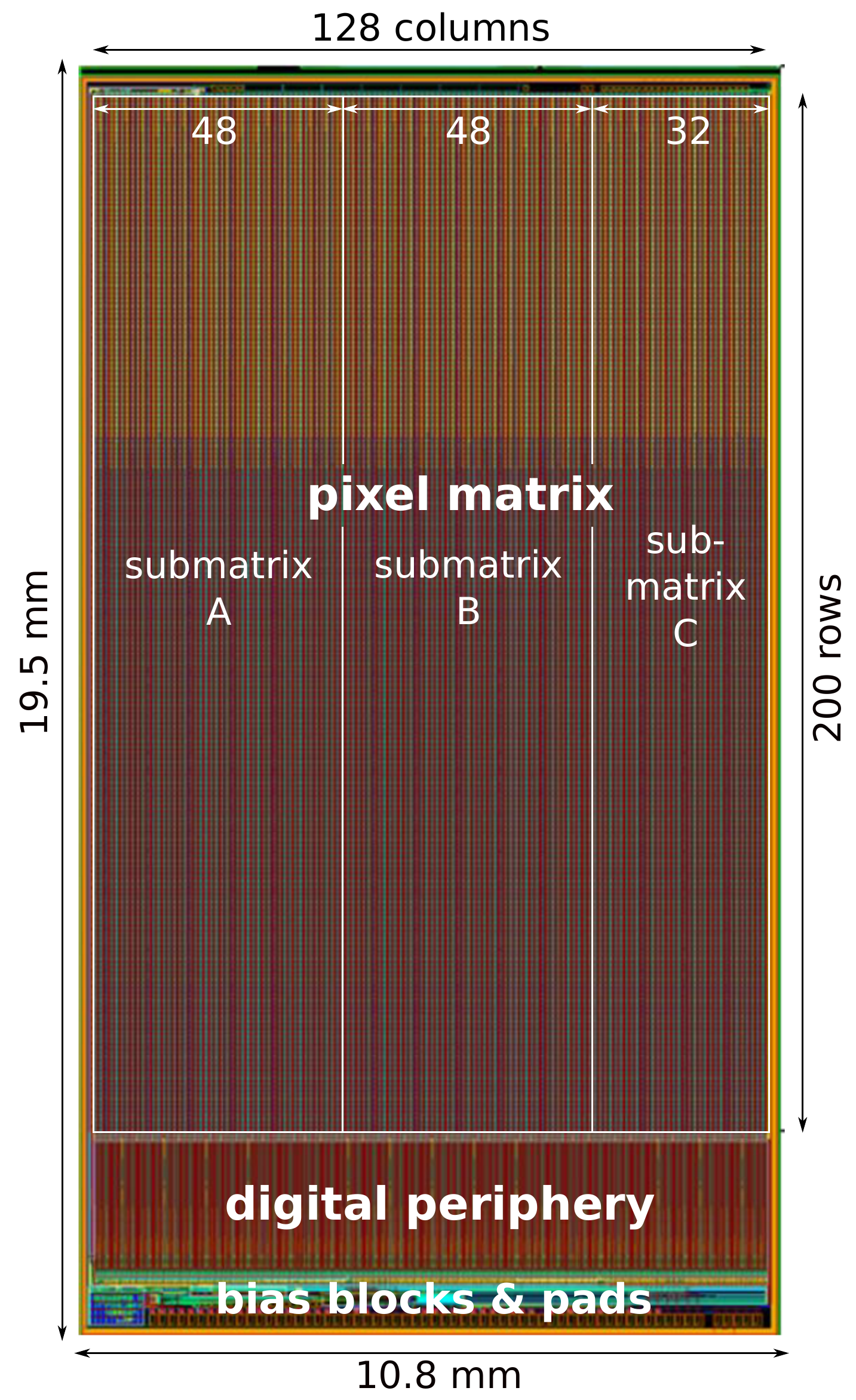}\label{fig3}}
\caption{The MuPix8 -- A large-area HV-MAPS chip.}
\end{figure}

The MuPix8 chip is produced in an AMS aH18 engineering run with a minimal gate length of $180\, \text{nm}$  together with further test chips. Table \ref{tab1} compares the design parameters of MuPix8 and MuPix7. The most notable change is the size of the prototype which enables the investigation of the column-drain architecture with the full column length of the final chip. Further the large pixel matrix of MuPix8 is segmented into three sub-matrices which are read out individually by three state machines, which are synthesized in one block, see \ref{fig3}.

The new prototype makes use of \(80\, \text{and}\, 200\,\Omega\text{cm}\) substrates, which increase the depletion zone by more than a factor 2 compared to previous prototypes which have been produced with the standard substrate of $20\,\Omega\text{cm}$ and thereby lead to a larger number of signal electrons per traversing charged particles. For $80\,\Omega\text{cm}$ and $-60\,\text{V}$ bias a depletion depth of $\approx 27\,\text{\textmu m}$ is expected. A further important change compared to previous chips is the use of enclosed transistors for analogue and biasing circuits which is expected to improve the robustness against total ionizing dose effects caused by irradiation, allowing for usage in harsher irradiation environments than Mu3e.

\begin{figure}[h]
\centering
\includegraphics[width = .90\textwidth]{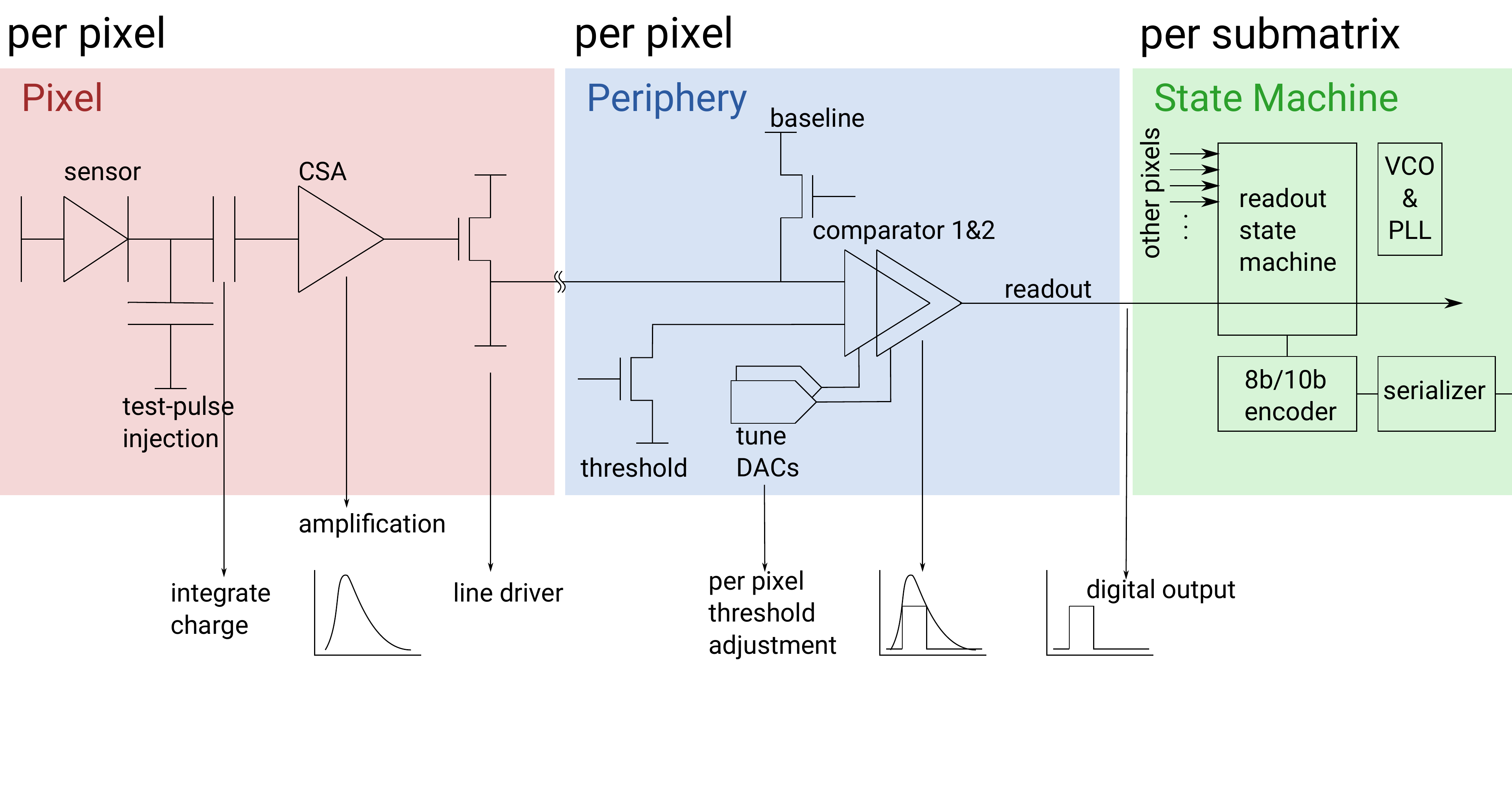}
\caption{A functional description of the MuPix architecture.}
\label{fig4}
\end{figure}

Following the MuPix-concept the chip structure is split into three functional parts as depicted in figure \ref{fig4}. Inside the active pixel matrix a charge sensitive amplifier and a signal line driver are implemented in the pixel diode. Each individual pixel has a point-to-point connection to its corresponding digital cell in the periphery. Here the signal is discriminated and a hit flag is stored along with a 10 bit timestamp. An additional 6 bit timestamp, which allows to measure the length of the analogue pulse, is implemented, see section \ref{sectwc}. The information is stored until it is read out by the state machine in a priority encoded scheme. The data is 8bit/10bit encoded, serialized and transmitted off-chip differentially with up to $1.6\,\text{Gbit}/\text{s}$.


\subsection{Analogue Pulse Transmission \& Crosstalk}
\label{sec:intro_xtalk}
\begin{figure}[h]
\centering
\includegraphics[width = .49\textwidth]{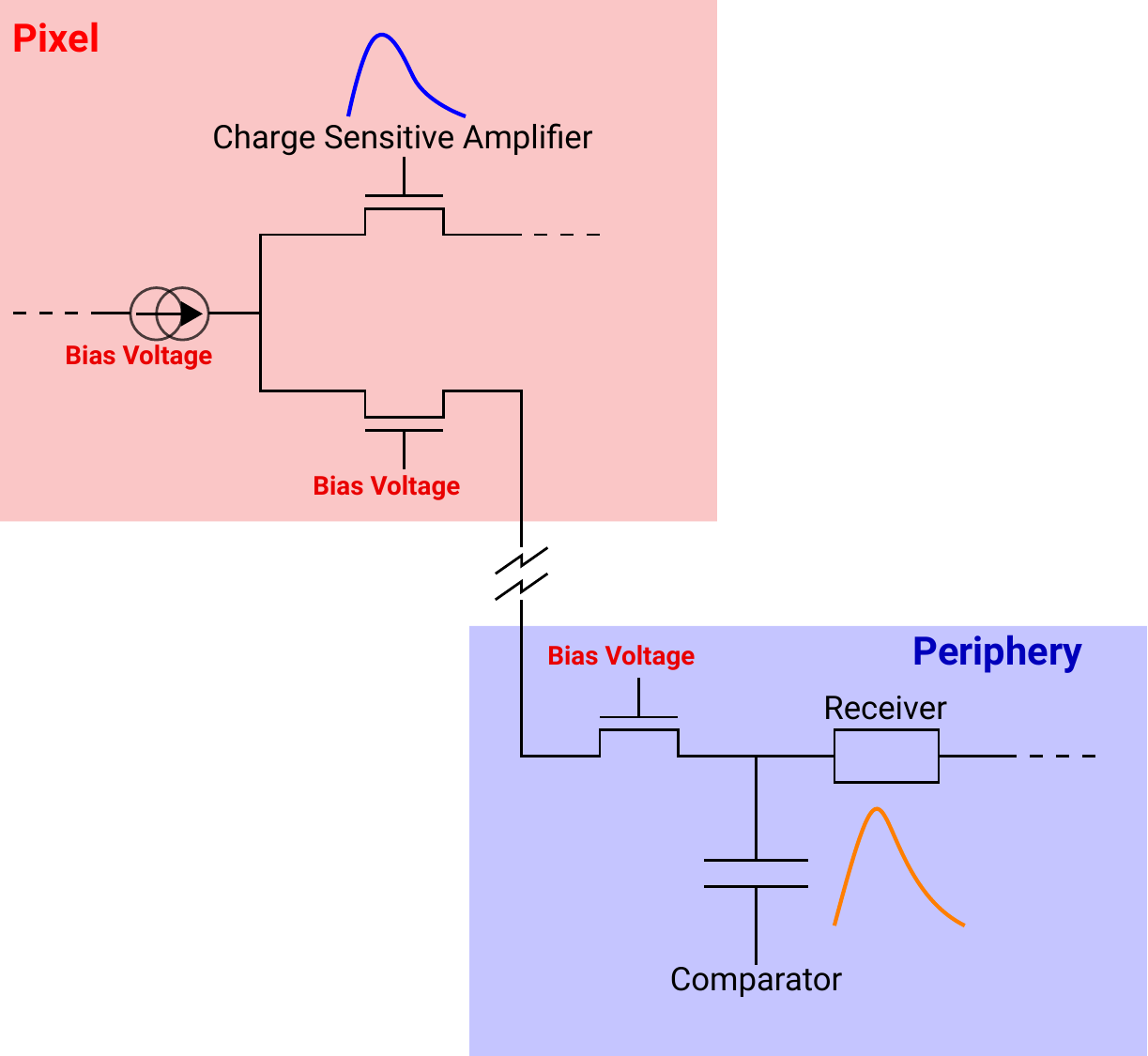}
\caption{The current driven signal transmission.}
\label{fig5}
\end{figure}

For the MuPix7, crosstalk was observed between the transmission lines from the in-pixel amplifier to the comparator in the periphery. It is explained by the spatial proximity of signal lines in the chips routing scheme creating a capacitive coupling~\cite{PIXEL:MuPix7}. For MuPix8, which is now exploring the full column length of \(2\,\text{cm}\) and an even higher routing density of the signal lines, this effect is expected to be more prominent if using a source follower as line driver, as implemented in MuPix7. To counteract this effect, a new current based transmission scheme, sketched in figure \ref{fig5}, is implemented which is expected to reduce the overall crosstalk as the voltage amplitude of the signal is small~\cite{Weber2016}. This new approach is implemented in sub-matrices B and C of MuPix8, while matrix A is making use of the MuPix7-like source follower.

\newpage
\subsection{Timewalk Correction}
\label{sectwc}
Timewalk was found to significantly contribute to the time resolution of MuPix7~\cite{Augustin:2018ppf}. For this reason different timewalk mitigation schemes are implemented in Mupix8. By measuring the pulse height or an equivalent value, the hit timestamp can be corrected for timewalk. The MuPix8 is the first MuPix prototype that samples a 6 bit pulse height information for each pixel-hit and allows for off-chip timestamp correction.

\begin{figure}[h]
\centering
\subfigure[ToT measurement with a voltage ramp.]{\includegraphics[height = .22\textheight]{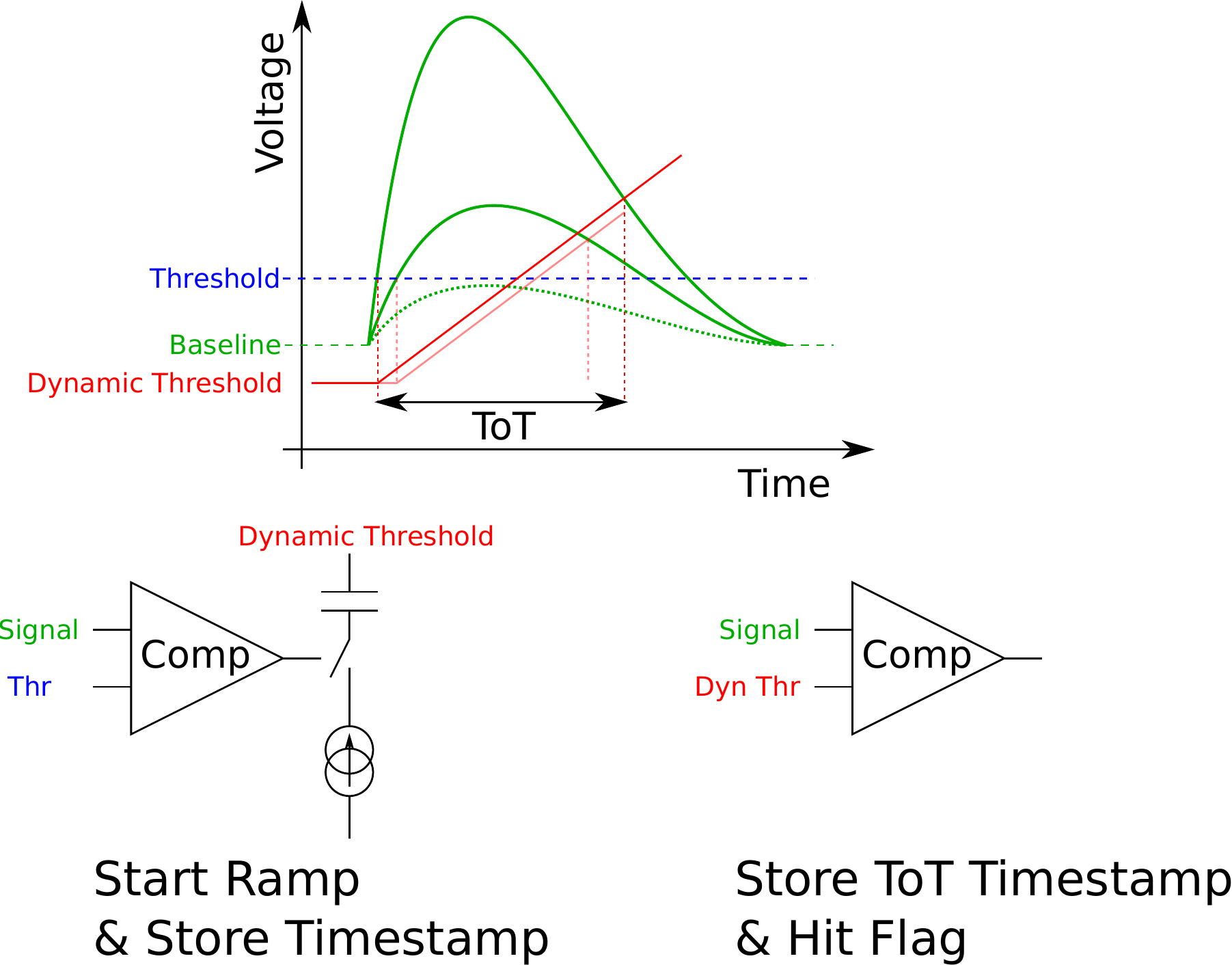}\label{fig6}}
\qquad
\subfigure[The two-threshold-approach.]{\includegraphics[height = .21\textheight]{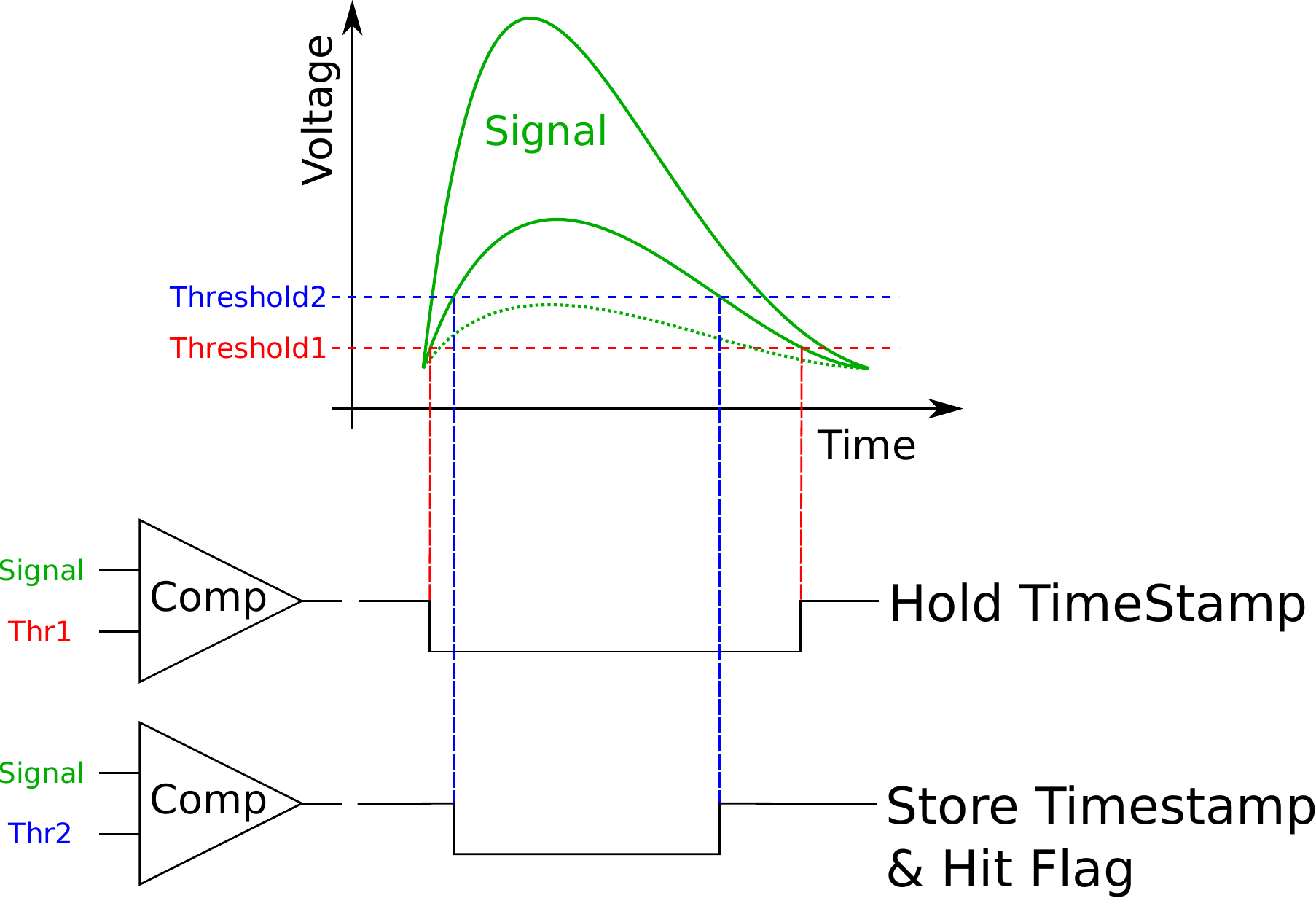}\label{fig7}}
\caption{Digital pixel cell: sampling modes.}
\end{figure}

The digital pixel cell contains two comparators with an individual threshold which can be used in three different modes. The simplest mode measures a Time-over-Threshold (ToT) for the discriminated pulse with a constant threshold. A more involved pulse height measurement is depicted in figure \ref{fig6} which makes use of a voltage ramp. The nearly perpendicular crossing angle between the ramp and signal pulse enables a ToT measurement with reduced jitter, as it is less effected by noise.

The third method presented in figure \ref{fig7} is not relying on the pulse height measurement, but is reducing timewalk effects directly on-chip. By choosing one threshold close to the baseline to sample the timestamp, the effect from varying rising edges is minimised. The second threshold is then used to discriminate against noise. Additionally also the ToT is sampled and can be used to further improve the time resolution.

The MuPix8 is used to investigate the performance of the different modes and to decide which is most suitable for the final sensor.

%
%
%

\section{Results}
\label{sec:results}
%
%
In the following key results from the laboratory and several testbeam campaigns are presented. The efficiency measurements have been performed with two different setups. Detailed scans have been performed with the MuPix-Telescope~\cite{Huth2018}. For sub-pixel studies the high precision of the Eudet-Telescope~\cite{EUDET} is utilized, using a $3\,\text{GeV}$ electron beam at the DESY test beam facility.

\subsection{Commissioning}
\label{sec:comm}
The first MuPix8 chips have been successfully commissioned in September 2017~\cite{Kroeger2017} and are fully functional. The chip is configured via the custom designed shift register with data rates up to \(6\,\text{Mbit}/\text{s}\) yielding a chip setup time of less than $100\,\text{ms}$. In a high-rate beam test at the MAMI facility the sub-matrices provided untriggered data streams with more than $10\,\text{MHits}/\text{s}$~\cite{Dittmeier2018}. Unexpectetly it is observed that the chip reaches its breakdown voltage at $-60\,\text{V}$, although it is designed for $-120\,\text{V}$. The early breakdown is still under investigation, however first observations hint towards fabrication issues~\cite{PIXEL:Breakdown}. A further novelty on the MuPix8 is the implementation of a pixel-switch additional to individual threshold tuning bits. The implementation was successful and allows to mask noisy pixel if necessary. All results in the following have been obtained using untuned chips, without pixel-masking if not indicated otherwise.

\subsection{Efficiency \& Noise Performance}
\label{sec:effi}
\begin{figure}[h]
\centering
\subfigure[No pixel-masking.]{\includegraphics[width=.44\textwidth]{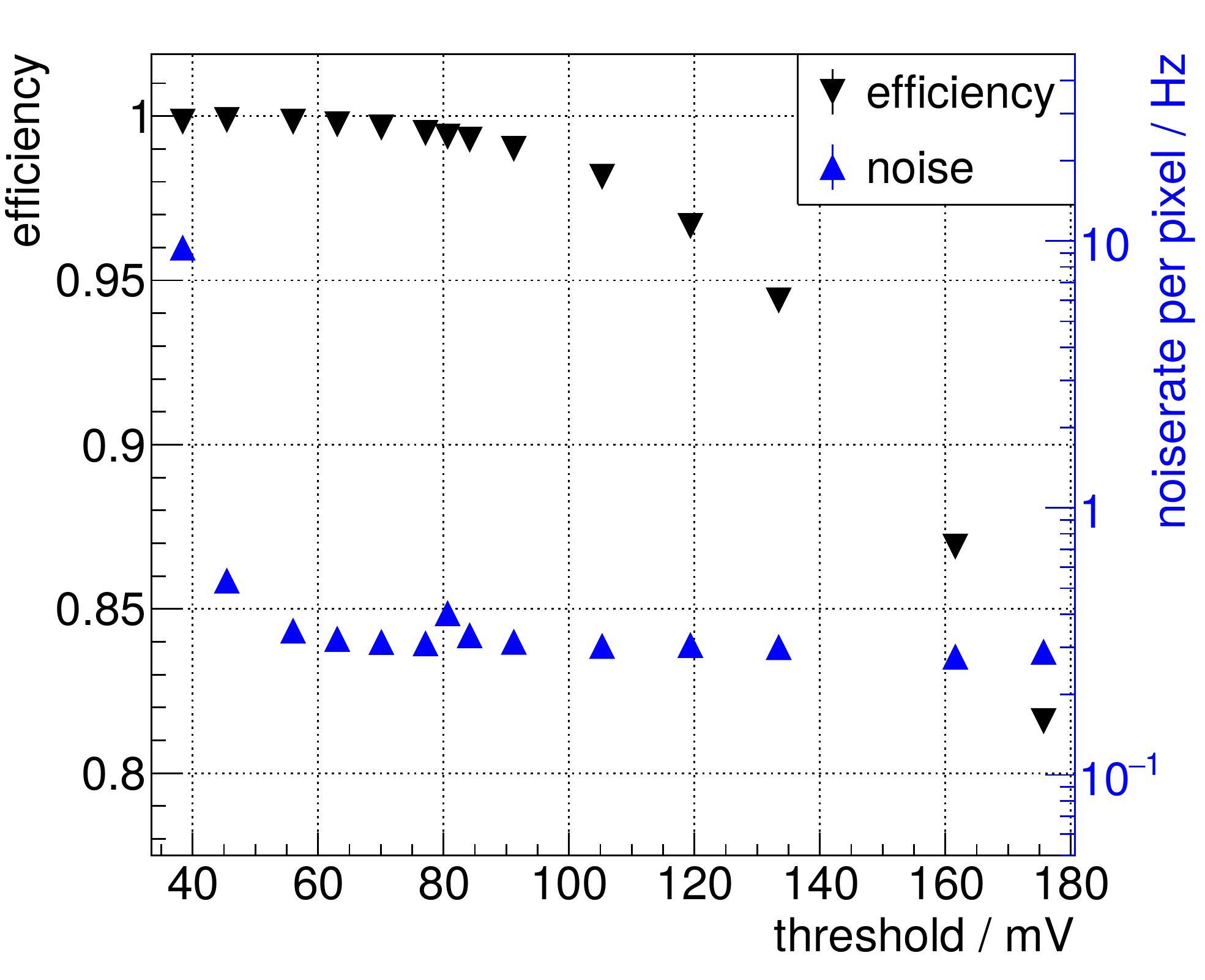}\label{fig8}}
\subfigure[Increased plateau with 25 masked pixels.]{\includegraphics[width=.44\textwidth]{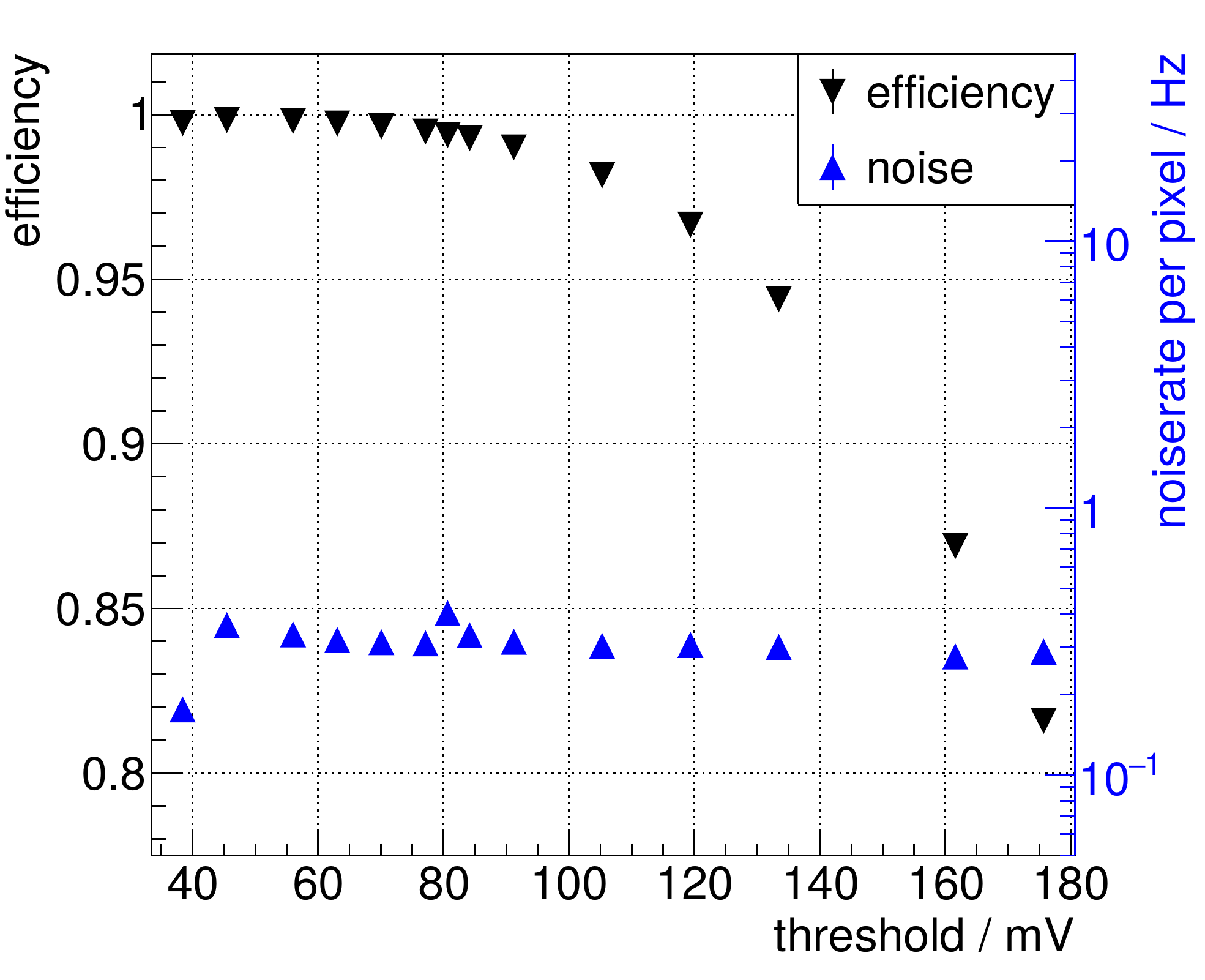}\label{fig9}}
\caption{MuPix8 efficiency and noise scan with the MuPix-telescope~\cite{Huth2018}.}
\label{fig:effi_scan}
\end{figure}

In figure \ref{fig:effi_scan} efficiency and noise are plotted as function of the detection threshold for an $80\,\Omega\text{cm}$ sensor at $-60\,\text{V}$ reverse bias. The scan in figure \ref{fig8} shows a $20\,\text{mV}$ wide plateau with an efficiency larger than $99.9\,\text{\%}$ at an average noise level below $1\,\text{Hz per pixel}$. The operation range is further increased with pixel-masking at the cost of reducing the efficiency by less than $2\,\text{\textperthousand}$, see figure \ref{fig9}. For this sensor $50\,\text{mV}$ threshold correspond to approximately $1200\,\text{electrons}$ with expected $3000\,\text{electrons}$ signal from a MIP. 

\begin{figure}[h]
\centering
\subfigure[$80\,\text{\&}\,200\,\Omega\text{cm}$ resistivities with $100\,\text{\&}\,725\,\text{\textmu m}$ sensor thickness.]{\includegraphics[width=.49\textwidth]{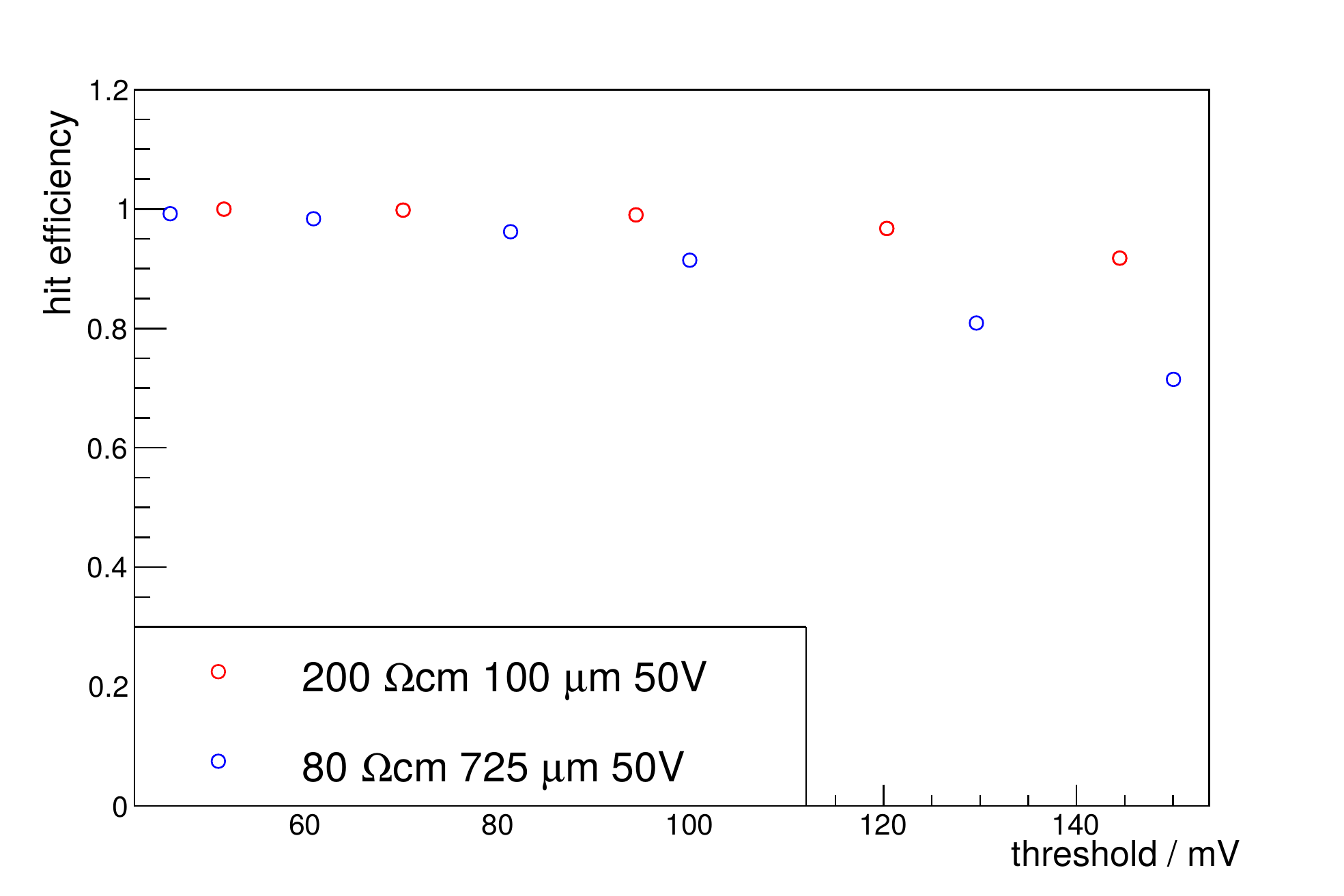}\label{fig10}}
\subfigure[Varying the reverse bias voltage.]{\includegraphics[width=.49\textwidth]{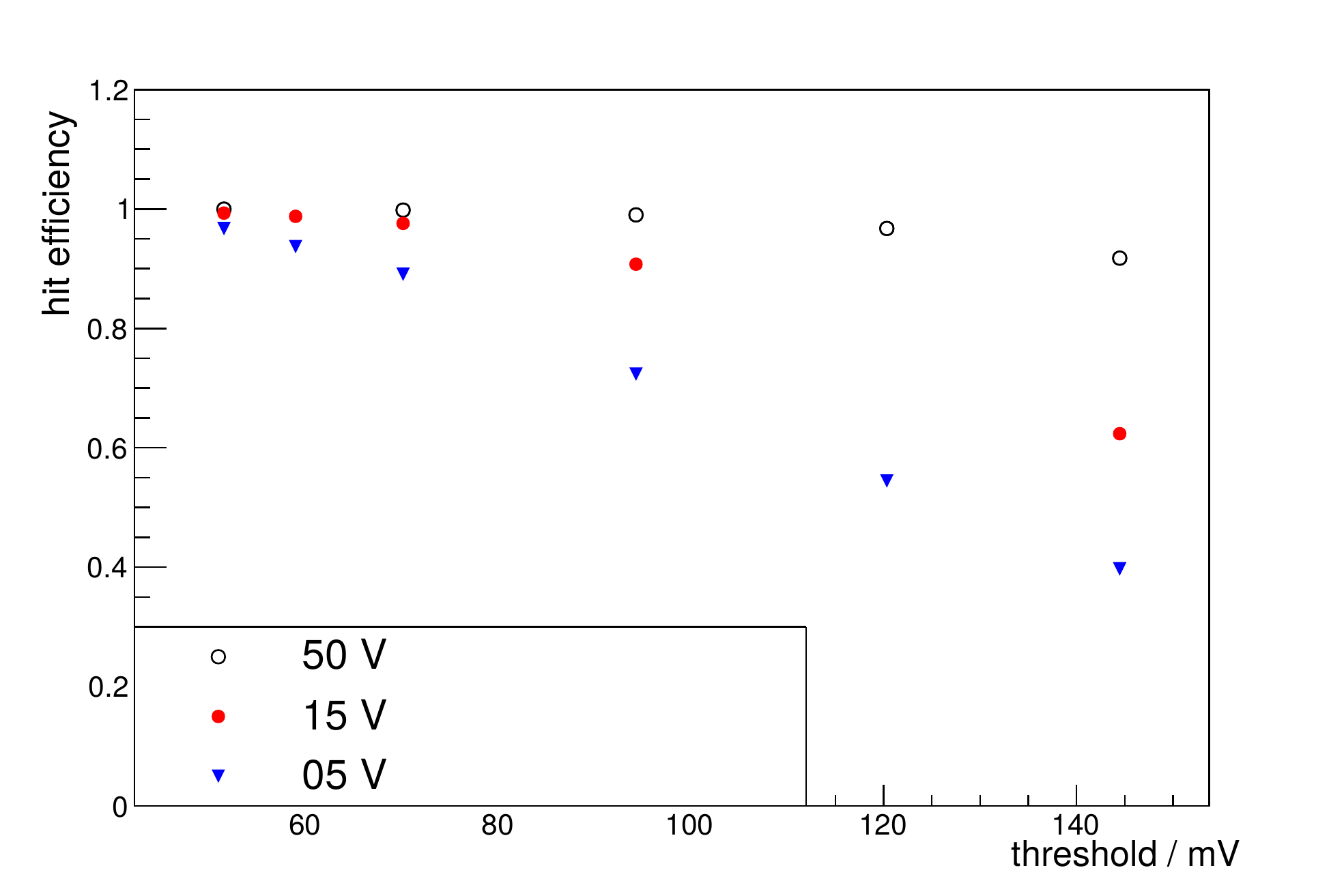}\label{fig11}}
\caption{Efficiency measurements for different threshold voltages, comparing resistivities and applied bias voltages for the MuPix8 sensor~\cite{Huth2018}.}
\label{fig:res_hv}
\end{figure}

With MuPix8, for the first time, different substrate resistivities and the effect on the thickness of the depletion region are investigated. Therefore threshold scans have been performed using two chips with $80\,\text{and}\,200\,\Omega\text{cm}$ resistivity respectively, biased to $-50\,\text{V}$, see figure \ref{fig10}. The scan reveals an even broader efficiency plateau for the $200\,\Omega\text{cm}$ chip. The bias voltage dependence is examined with threshold scans of the $200\,\Omega\text{cm}$ chip for $-5$, $-15$ and $-50\,\text{V}$, see figure \ref{fig11}. Here an overall increase of the efficiency and the manifestation of the high efficiency plateau is observed for increasing reverse bias. These measurements agree perfectly with the expectation, as for higher resistivity $\rho$ and a larger reverse bias U the depletion zone grows proportional to $\sqrt{\rho\cdot\text{U}}$, creating more signal electrons per penetrating charged particle. By varying the bias voltage from $-5\ \text{to}\ -50\,\text{V}$ one expects a growth of the depletion zone by more than a factor 3 and further achieves a full lateral depletion in-between the pixels.

\begin{figure}[h]
\centering
\subfigure[Settings:\newline $50\,\text{mV}$ threshold, $-60\,\text{V}$ bias voltage.]{\includegraphics[width=.45\textwidth]{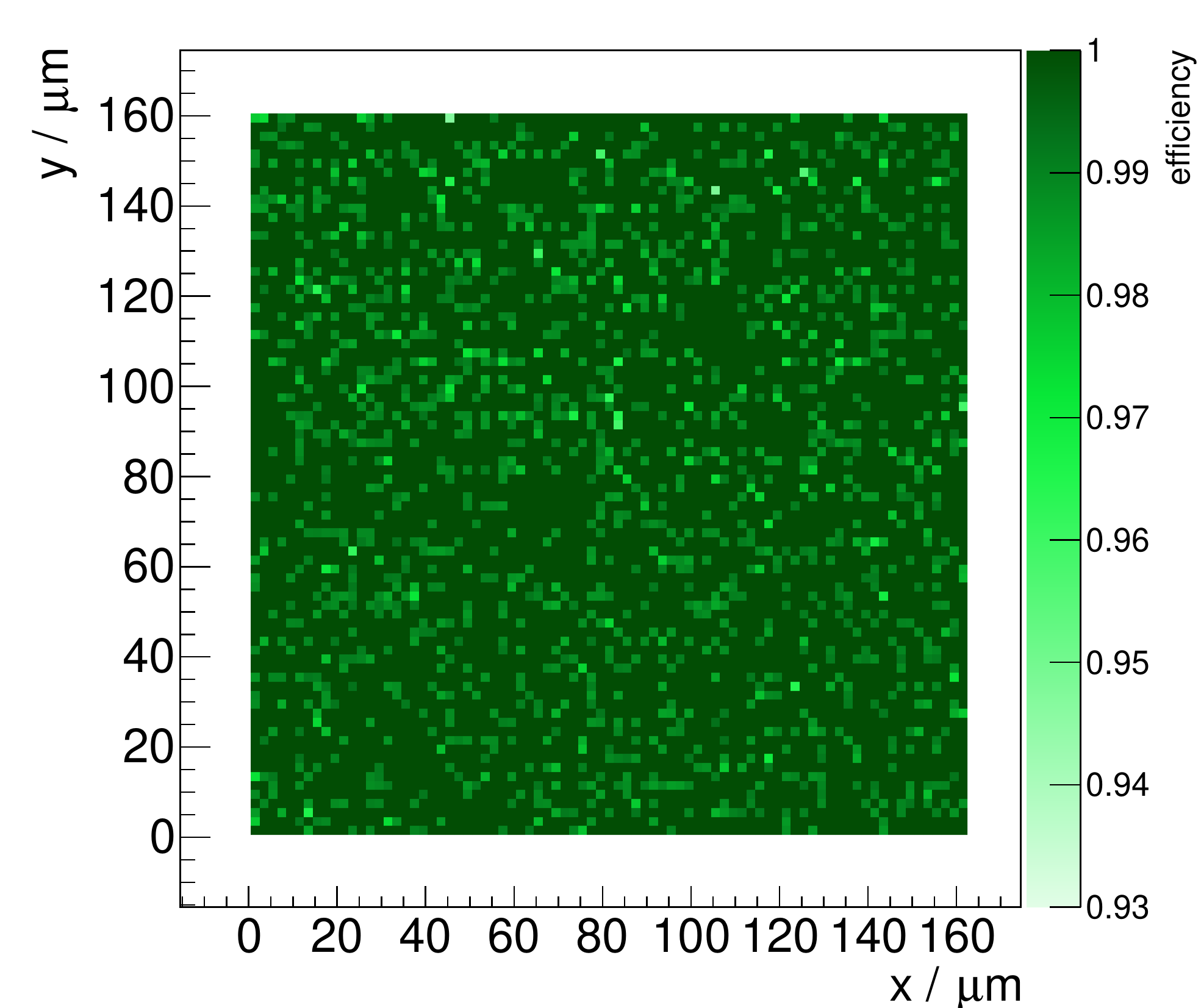}\label{fig12}}
\subfigure[Settings:\newline $50\,\text{mV}$ threshold, $-15\,\text{V}$ bias voltage.]{\includegraphics[width=.45\textwidth]{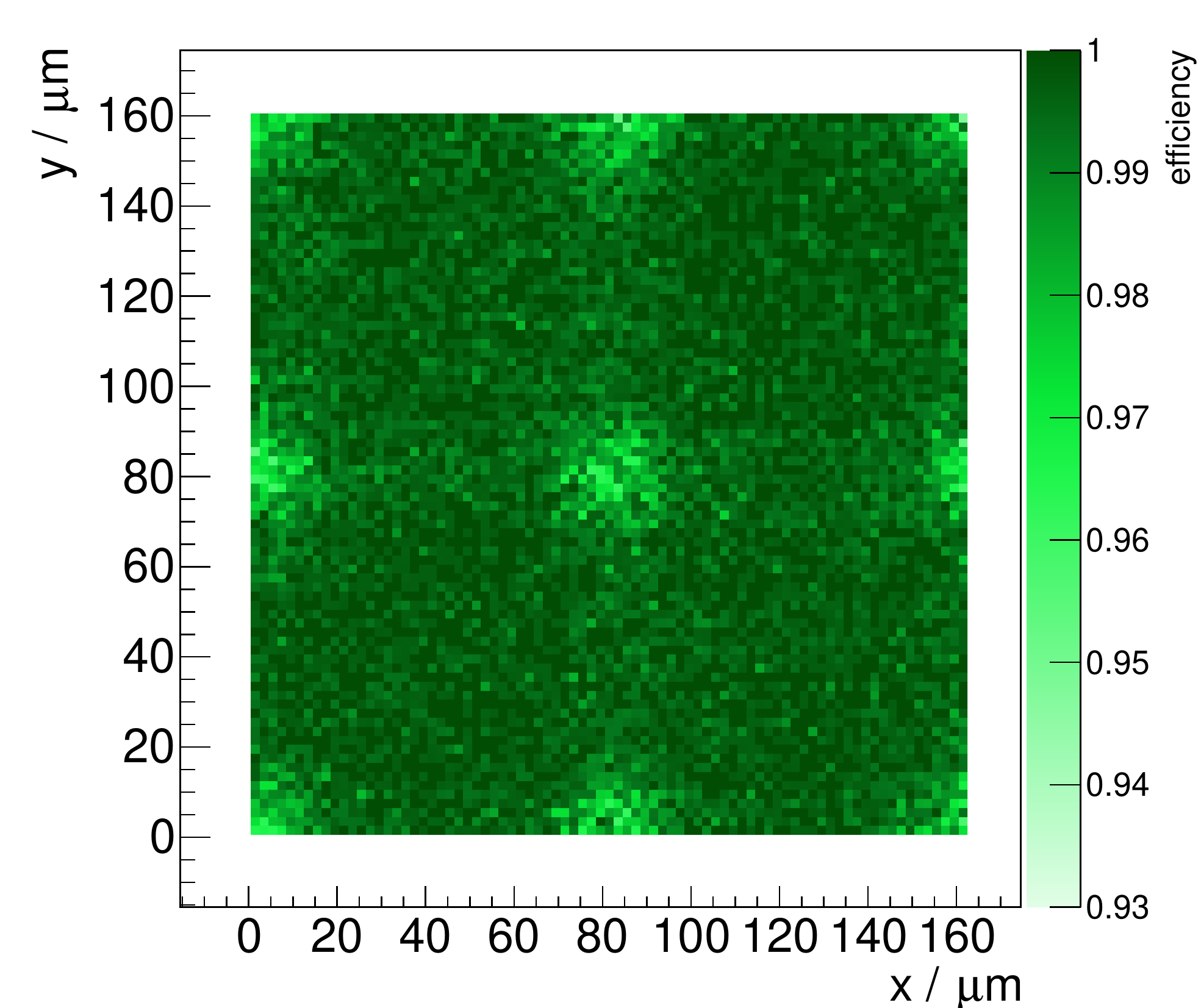}\label{fig13}}
\caption{Studies of sub-pixel efficiencies in a \(2\times2\) pixels folded view~\cite{Huth2018}.}
\label{fig:sub-pixel}
\end{figure}

To further confirm the full lateral depletion a sub-pixel analysis is performed with the help of the high resolution Eudet-Telescope~\cite{EUDET}. The results are shown in figure \ref{fig:sub-pixel}, folding the full matrix to $2\times2$ pixels for improved statistics. For a working point in the plateau, figure \ref{fig12}, no inefficiency is expected and no sub-pixel structures are recognisable. To observe sub-pixel effects a working point with reduced efficiency needs to be chosen. To achieve this the bias voltage is reduced, shrinking the depletion zone and therefore the signal charge. With these settings inefficiencies are observed in the pixel corners. Here the signal charge created by perpendicular tracks is shared by up to 4 pixels, reducing the effective signal per pixel.

\newpage
\subsection{Time Resolution}
\label{sec:timeres}
The time resolution of MuPix8 is studied with a $^{90}$Sr source and a coincidence setup, consisting of a MuPix8 chip and a scintillator. By measuring the time difference between the hit on the MuPix and the scintillator the time resolution is determined~\cite{Hammerich2018}. For the measurement of the deposited energy the 6 bit ToT measurement of the digital pixel cell is utilized.

\begin{figure}[h]
\centering
\subfigure[Measurement of the column and row dependent delay in $8\,\text{ns}$ bins.]{\includegraphics[width=.48\textwidth]{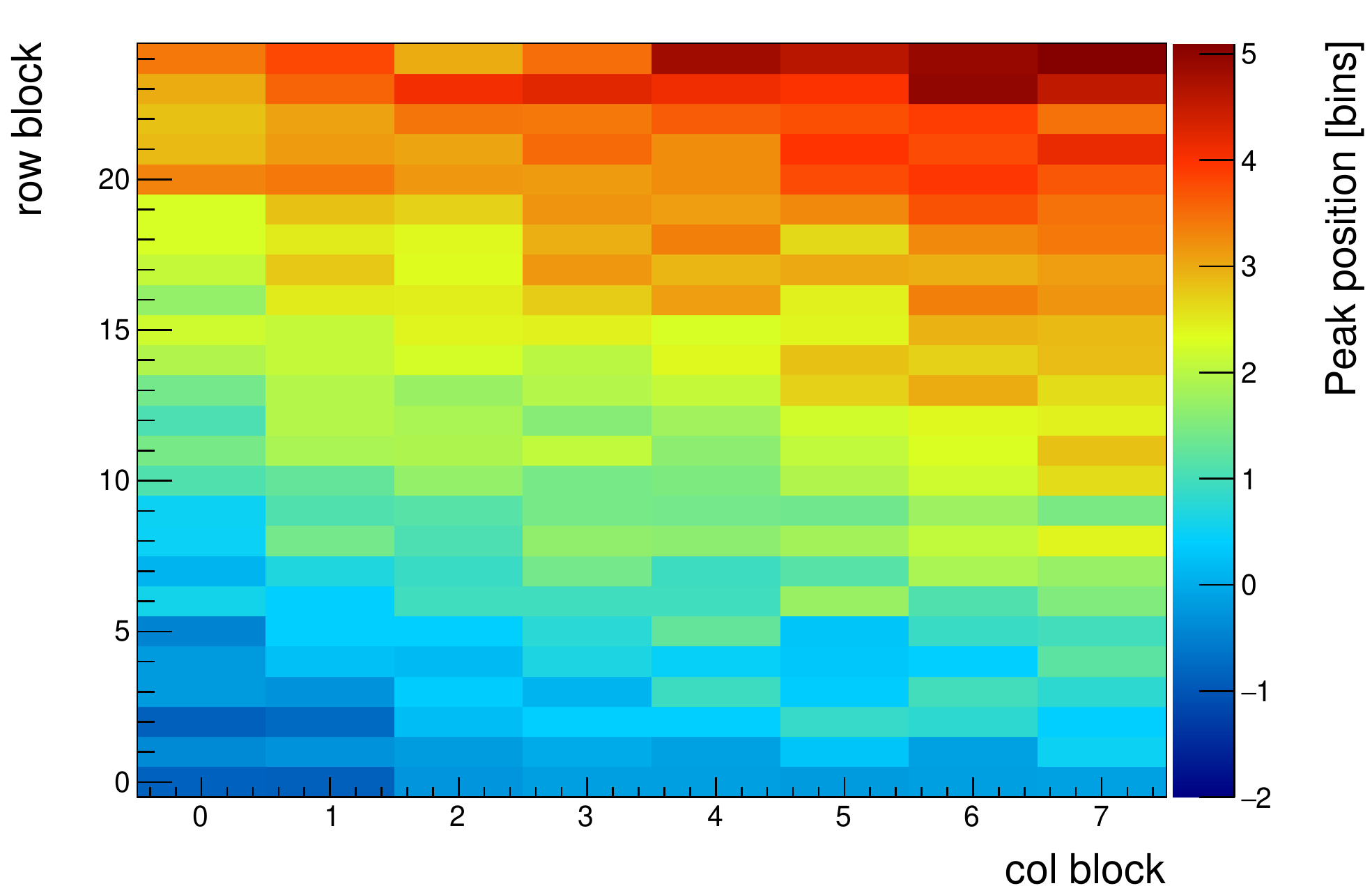}\label{fig15}}
\subfigure[Timewalk after delay correction.]{\includegraphics[width=.48\textwidth]{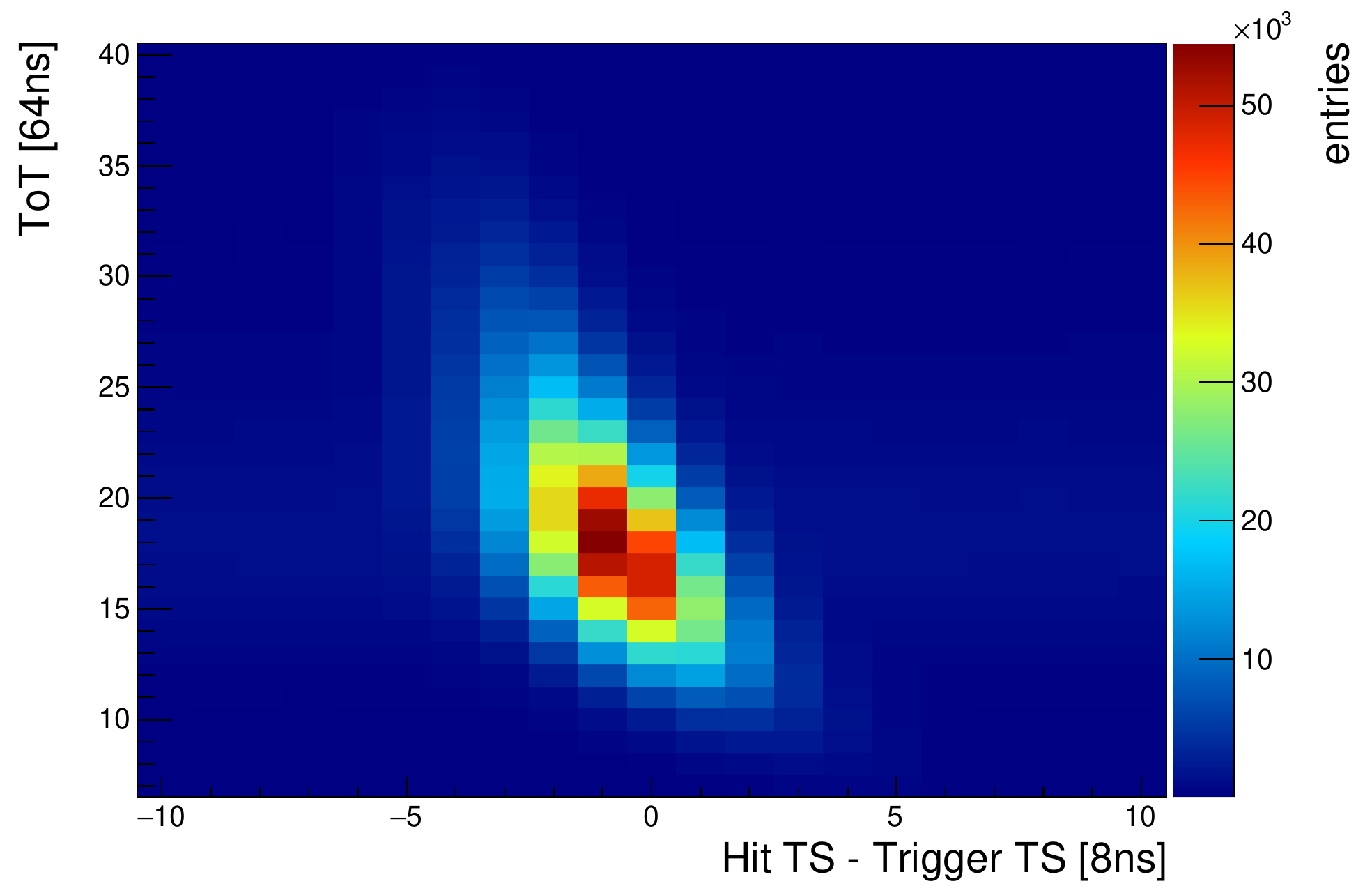}\label{fig16}}
\subfigure[Time resolution improvement after pixel delay and timewalk correction.]{\includegraphics[width=.63\textwidth]{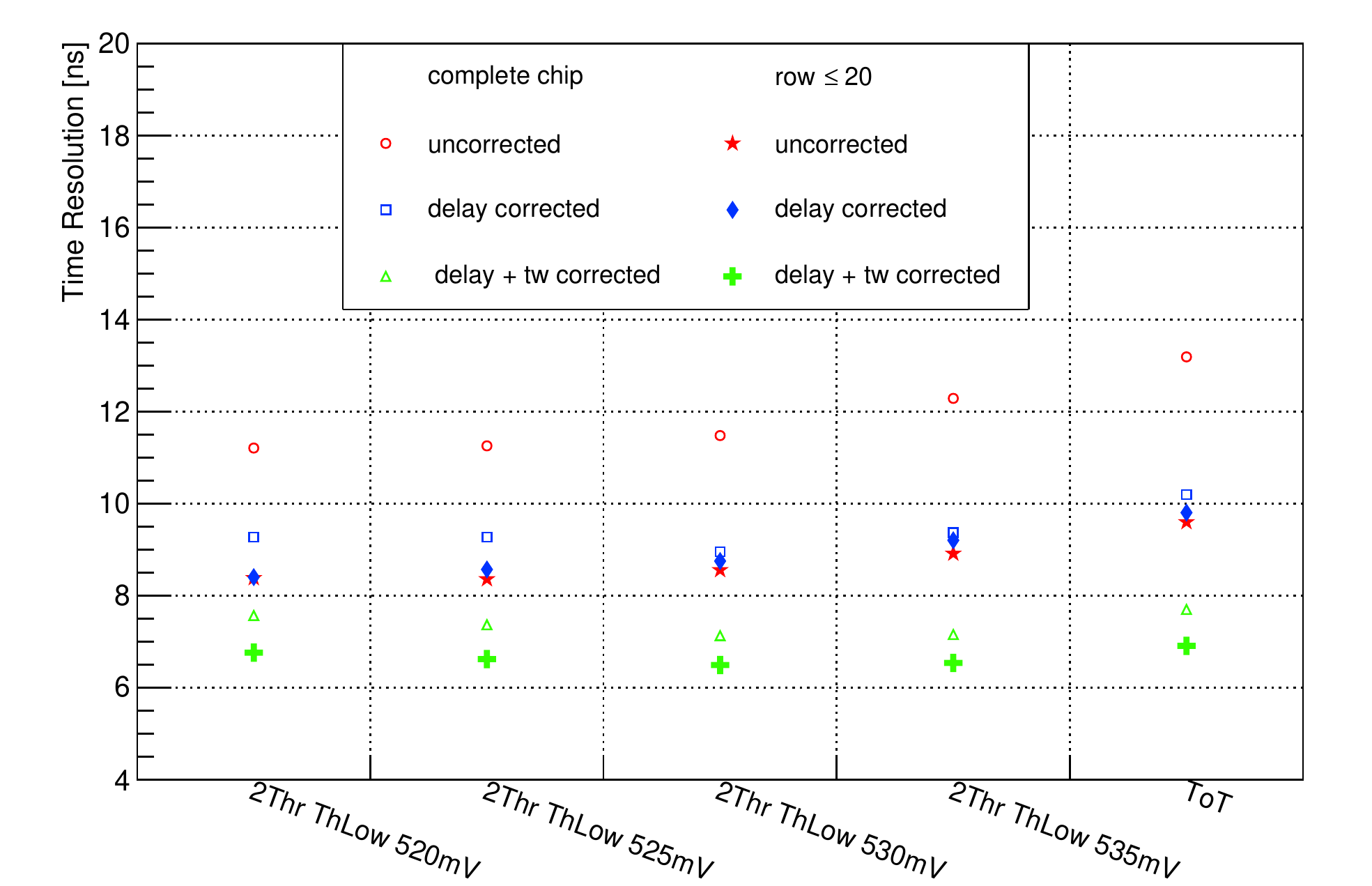}\label{fig17}}
\caption{Results of time resolution studies~\cite{Hammerich2018}.}
\end{figure}

Without any additional processing a time resolution of $\sigma = 14\,\text{ns}$ is achieved for the chip, averaged over all pixels. This already fullfills the Mu3e specification of $\sigma < 20\,\text{ns}$, but is worse than the result obtained with the predecessor~\cite{PIXEL:MuPix7}. Two effects are mainly responsible: firstly timewalk from variations of the rising signal edge, depending on the amount of deposited energy and secondly row dependent variations of the signal delays caused by different line lengths of the pixel point-to-point connections and thus signal line capacitances. The pixel position dependent delay is measured, see figure \ref{fig15}, and shows a dependence in both row and column direction. While the row dependence is explained by the increasing line length, the column change is interpreted as the result of a voltage drop along the column. The data of $6\times4$ pixels has been combined in blocks to enhance statistics and is fitted with a two-dimensional function which is used to correct the column and row dependence for every pixel. After applying the delay correction the time resolution is improved and the timewalk becomes visible, see figure \ref{fig16}. For the timewalk correction the peak position of the pixel-scintillator time difference is determined for individual ToT bins and shifted to zero. With this method a time resolutions of $\sigma = 6.5\,\text{ns}$ has been achieved.

All above achieved improvements require an off-chip calculation. The two-threshold approach implemented in MuPix8, see section  \ref{sectwc}, aims for an on-chip suppression of timewalk effects. With this method a $1\,\text{to}\, 2\,\text{ns}$ improvement is achieved compared to the single threshold implementation, see figure \ref{fig17}. The point denoted with ToT represents the single-threshold scheme, for the two-threshold method different lower threshold settings (ThLow) have been evaluated. Further the plot includes results obtained for the first 20 rows only, which have the shortest line lengths and also a reduced length spread compared to the rest of the chip. For these pixels the obtained time resolutions are significantly better, in particular if no corrections are applied.

\subsection{Crosstalk}
\label{sec:xtalk}
\begin{figure}[h]
\centering
\includegraphics[width=.6\textwidth]{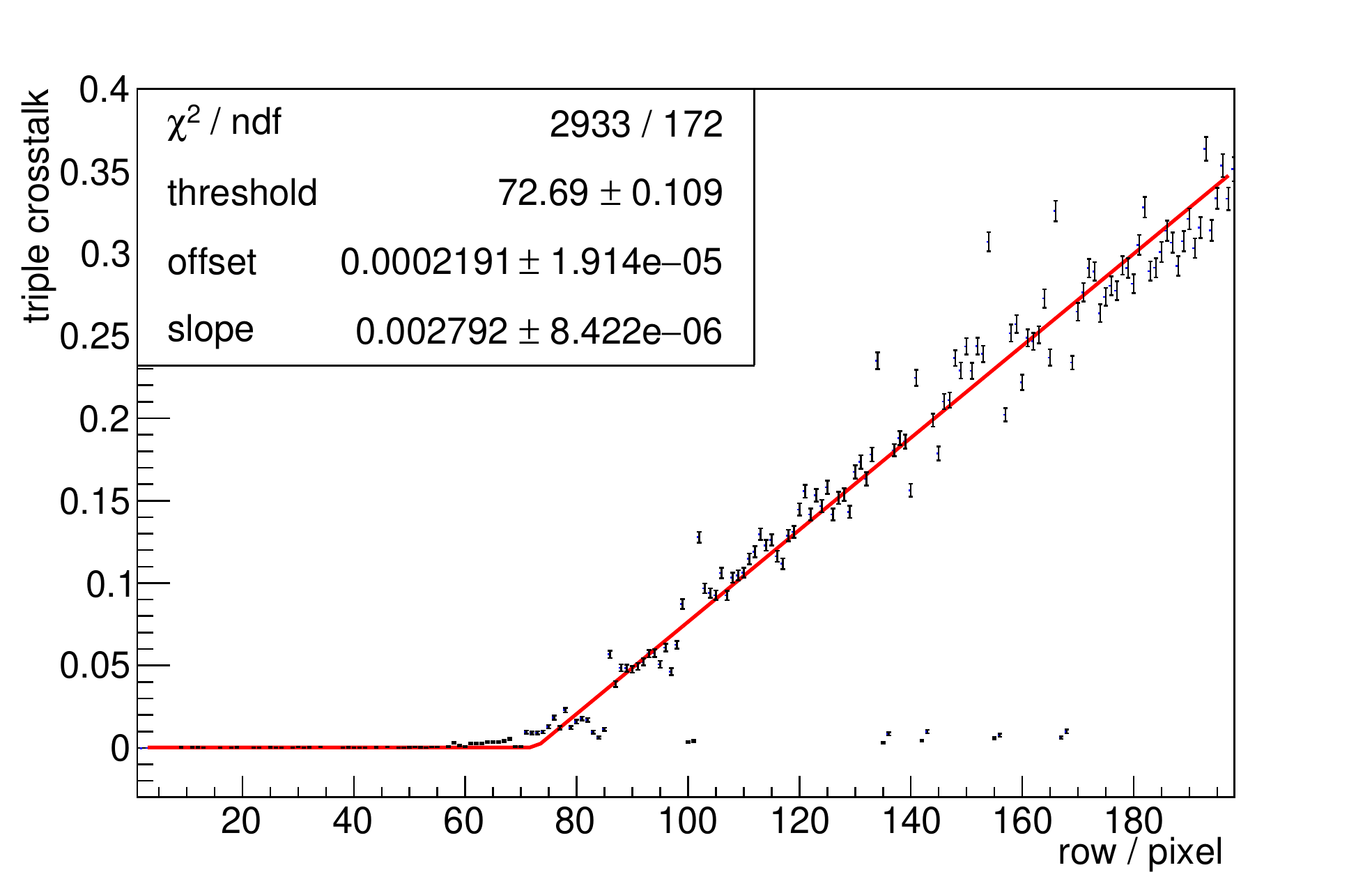}
\caption{The row dependence of triple crosstalk for sub-matrix A using the source follower as line driver~\cite{Huth2018}.}
\label{fig19}
\end{figure}

As expected, crosstalk is observed for the source follower approach implemented in MuPix8 sub-matrix A. An example is shown in figure \ref{fig19} which shows the triple crosstalk probability depending on the row position of the pixel. The hit topology to study triple crosstalk is three hits in neighboring pixels in the same column. This structure is a very distinct and also a very clean method to detect crosstalk if particles enter with vertical impact. The data show a linear increase of the triple crosstalk probablity and thus the inter-line capacity which scales with the signal line length and the row number accordingly. This clearly shows the disadvantage of the source follower as line driver approach, as for long lines the hit multiplicity and thus the load for the readout is increased by the additional hits.

An alternative is the current driven signaling scheme implemented in sub-matrices B\&C. The first results are very promising as they yield an efficiency of $>99\,\text{\%}$ at a time resolution of $\sigma < 15\,\text{ns}$ after corrections. More over, it successfully solves the crosstalk issue: no significant crosstalk is observed for theses sub-matrices. The performance of the current driver is expected to improve further in the next submission as the current implementation has a known issue in the biasing.

\section{Summary and Outlook}
\label{sec:sum}
The MuPix8 chip shows an excellent performance and proves the successful scaling of the HV-MAPS technology. Further, the use of higher ohmic substrates of up to \(200\,\Omega\text{cm}\) increases the amount of collectable primary charges, creating a broad working plateau with efficiencies $>99.9\,\text{\%}$. Using the pixel's additional 6-bit ToT information allows to perform a timewalk correction yielding an excellent time resolution of $6.5\,\text{ns}$.

With MuPix8 also the crosstalk issue is addressed. The predicted crosstalk enhancement compared to MuPix7 for the source follower scheme, due to longer and denser point-to-point connections is observed. An alternative is the implementation of the current driven signaling scheme, which proved to prevent crosstalk entirely. First measurements show that the performance of this approach yields a worse time resolution, however, the reason for this is found in the circuitry and can be improved for the next submission.

The measured time resolution of $\mathcal{O}(5\,\text{ns})$, which fulfills the Mu3e requirement of $\sigma<20\,\text{ns} $, can only be achieved with offline corrections. For the untriggered readout structure of the Mu3e experiment this would demand an additional involved correction step in the data flow, which is undesirable. The two-threshold approach was successfully implemented and reduced the timewalk effect. Furthermore the pixel position dependent delay can be avoided by using the same line length for all pixels.

Based on theses results and conclusions the next submission of the large-scale chip MuPix10 is already in preparation. MuPix10 will have the final chip size required for the Mu3e experiment with a $2\times2\,\text{cm}^2$ active pixel matrix and is going to be used to build the first module prototypes for the Mu3e pixel detector.


\acknowledgments

H.~Augustin, A.~Herkert, A.~Meneses~Gonzales and A.~Weber acknowledge support by the HighRR research training group [GRK 2058].
S.~Dittmeier and L.~Huth acknowledge support by the Internationale Max Planck Research School for Presicion Tests of Fundamental Symmetries.

This work has been supported by the Cluster of Excellence ``Precision Physics, Fundamental Interactions, and Structure of Matter'' (PRISMA$^+$ EXC 2118/1) funded by the German Research Foundation (DFG) within the German Excellence Strategy (Project ID 39083149).

Measurements leading to these results have been performed at the Test Beam 
Facility at DESY Hamburg (Germany), a member of the Helmholtz Association (HGF). We would like to thank the coordinators and support at DESY for the excellent test beam environment.

Further we would like to thank the PSI and the MAMI facility in Mainz for providing high rate test beams under excellent conditions.



\bibliographystyle{unsrt_collab_comma}
\bibliography{mybib}
\end{document}